\documentclass[twocolumn,showpacs,preprintnumbers,amsmath,amssymb,prl]{revtex4}

\usepackage{graphicx}
\usepackage{dcolumn}
\usepackage{bm}
\usepackage{color}
\begin{document}

\bibliographystyle{naturemag}

\preprint{}

\title{Magnetic field and temperature sensing with atomic-scale spin defects in silicon carbide}

\author{H.~Kraus$^{1}$}
\author{V.~A.~Soltamov$^{2}$}
\author{F.~Fuchs$^{1}$}
\author{D.~Simin$^{1}$}
\author{A.~Sperlich$^{1}$}
\author{P.~G.~Baranov$^{2}$}
\author{G.~V.~Astakhov$^{1}$} 
\email[E-mail:~]{astakhov@physik.uni-wuerzburg.de}
\author{V.~Dyakonov$^{1,3}$}
\email[E-mail:~]{dyakonov@physik.uni-wuerzburg.de}

\affiliation{$^1$Experimental Physics VI, Julius-Maximilian University of W\"{u}rzburg, 97074 W\"{u}rzburg, Germany \\
$^2$Ioffe Physical-Technical Institute, 194021 St.~Petersburg, Russia\\ 
$^3$Bavarian Center for Applied Energy Research (ZAE Bayern), 97074 W\"{u}rzburg, Germany}

\begin{abstract}
Quantum systems can provide outstanding performance in various sensing applications, ranging from  bioscience to nanotechnology. Atomic-scale  defects in silicon carbide are very attractive in this respect because of the technological advantages of this material and favorable optical and radio frequency spectral ranges to control these defects. We identified several, separately addressable spin-3/2 centers in the same silicon carbide crystal, which are immune to nonaxial strain fluctuations. Some of them are characterized by nearly temperature independent axial crystal fields, making these centers very attractive for vector magnetometry. Contrarily, the zero-field splitting of another center exhibits a giant thermal shift of $- 1.1 \, \mathrm{MHz / K}$ at room temperature, which can be used for thermometry applications. We also discuss a synchronized composite clock exploiting spin centers with different thermal response.  
\end{abstract}

\date{\today}


\maketitle

Nanoscale magnetic field and temperature sensing based on quantum properties of the nitrogen-vacancy (NV) defect in diamond \cite{Gruber:1997gk} has been proposed  \cite{Chernobrod:2005jb, Degen:2008jh, Taylor:2008cp, Toyli:2012gl}  and  demonstrated at ambient conditions \cite{Maze:2008cs, Balasubramanian:2008cz, Toyli:2013cn}. This sensing technique can potentially allow for monitoring neuron activity \cite{Hall:2012fe}, imaging of single proton spins in complex molecular systems \cite{Staudacher:2013kn, Mamin:2013eu} and measuring the heat produced by chemical reactions inside living cells \cite{Neumann:2013hc, Kucsko:2013gq}. The principle of sensing is based on the optical readout of the spin resonance frequency subject to the crystal field $D$ 
and magnetic field $B$. 

These successful experiments stimulated the search for other solid-state systems with similar properties. Atomic-scale defects in silicon carbide (SiC) are attractive in this respect because they have complementary abilities to the NV defect in diamond. Apart from the obvious technological opportunities due to the well-developed device fabrication protocols in SiC, (i) these defects are optically active in the near infrared, characterised by a deep tissue penetration; and (ii) they can be controlled by a radio frequency (RF) field in the MHz spectral range, used in standard magnetic resonance imaging systems. Because about 250 SiC polytypes are known, there should exist more than thousand different spin defects in SiC with distinct characteristics \cite{Falk:2013jq, Kraus:2013di}. One can select a defect with the most suitable properties for a concrete task, which is not possible for one universal sensor. 

Here, we identify separately addressable spin centers within the primary intrinsic defects in 6H-SiC and 4H-SiC. Some of these half-integer spin ($S = 3/2$) centers are insensitive to nonaxial strain as a consequence of the Kramers degeneracy theorem and reveal within our accuracy temperature independent crystal field. Thermal/strain fluctuations may limit the magnetometry performance. In particular, thermal fluctuations lead to fluctuations of the spin resonance frequency because $D$ is generally a function of temperature $(T)$. The NV defect demonstrates the thermal shift of the spin resonance frequency $\beta =- 74 \, \mathrm{kHz / K}$  \cite{Acosta:2010fq}, and temperature fluctuations of $1 \, \mathrm{^{\circ} C}$ results in a magnetic field uncertainty of a few $\mathrm{\mu T}$. By stabilizing temperature and using advanced measurement protocols in a bias magnetic field, this uncertainty can be significantly reduced \cite{Fang:2013dw}, while in case of SiC this is already the intrinsic property. 

 On the other hand, a large thermal shift is necessary for temperature sensing. We also identify spin-3/2 centers, demonstrating giant thermal shifts up to  $\beta = - 1.1 \, \mathrm{MHz / K}$ at ambient conditions. This is 14 times larger than that for the NV defect and hence can potentially be used to enhance the thermometry sensitivity.

\section{Room-temperature ODMR of intrinsic defects  in 6H-SiC}
 
\begin{figure}[btp]
\includegraphics[width=0.99\columnwidth]{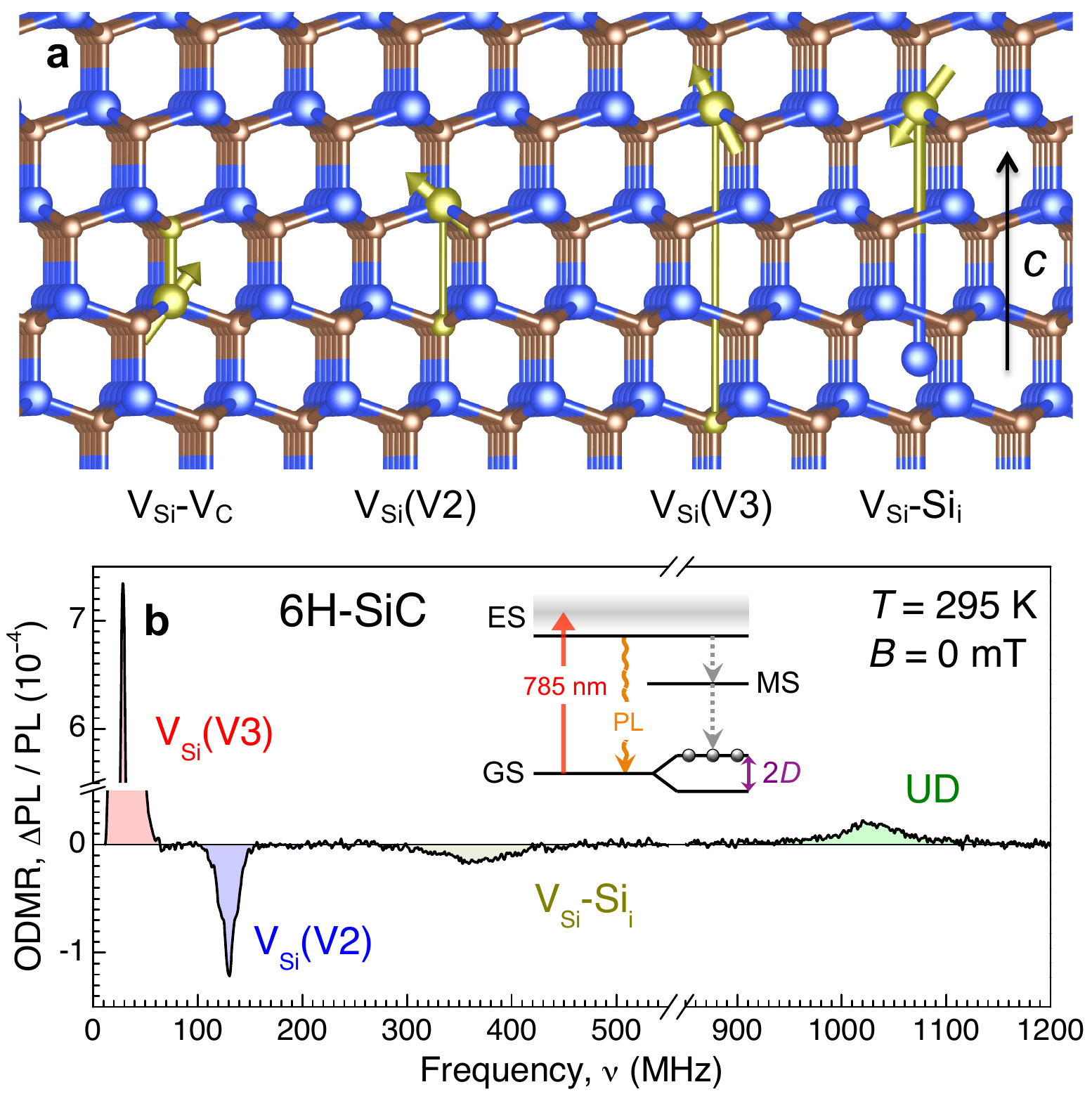}
\caption{Optical detection of spin defects in 6H-SiC. (a) Crystallographic structures of the Schottky [$\mathrm{V_{Si}}$-$\mathrm{V_{C}}$, $\mathrm{V_{Si}(V2)}$, $\mathrm{V_{Si}(V3)}$] and Frenkel ($\mathrm{V_{Si}}$-$\mathrm{Si_{i}}$) defects. All the defects have the symmetry axis oriented along the $c$-axis of the crystal lattice. (b) Room-temperature zero-field ODMR spectrum. Inset:  optical pumping scheme of spin defects. } \label{fig1}
\end{figure}

The Schottky and Frenkel defects are the primary stoichiometric  defects in solids. A divacancy ($\mathrm{V_{Si}}$-$\mathrm{V_{C}}$)---consisting of chemically bound silicon vacancy ($\mathrm{V_{Si}}$) and carbon vacancy ($\mathrm{V_{C}}$)---is an example of a Schottky defect in SiC [Fig.~\ref{fig1}(a)] being extensively investigated \cite{Vainer:1981vj, Baranov:2005em, Son:2006im, Koehl:2011fv}. Another example of a Schottky defect is an isolated silicon vacancy \cite{Vainer:1981vj, Wimbauer:1997fj, Sorman:2000ij, Mizuochi:2002kl, Baranov:2011ib, Soltamov:2012ey, Riedel:2012jq, Fuchs:2013dz}   perturbed by the nearest carbon vacancy along the $c$-axis of the SiC lattice \cite{Kraus:2013di}. There are several possible configurations and two of them are labeled in Fig.~\ref{fig1}(a) as  $\mathrm{V_{Si}(V2)}$ and $\mathrm{V_{Si}(V3)}$. Frenkel defects have also been identified in SiC and one of such defects---a silicon vacancy and a corresponding interstitial Si atom located at a distance of 6.5~{\AA} along the $c$-axis ($\mathrm{V_{Si}}$-$\mathrm{Si_{i}}$) \cite{vonBardeleben:2000jg}---is shown in Fig.~\ref{fig1}(a). 

A common feature of these defects in SiC is that they have a high-spin ground state  \cite{Baranov:2005em, vonBardeleben:2000jg}.  The corresponding spin Hamiltonian is written in the form
\begin{equation}
\mathcal{H} = g_e \mu_B   \mathbf{B}  \mathbf{S} + D(T) [ S_z^2 - S(S+1) / 3] \,.
 \label{Hamiltonian}
\end{equation}
Here, $g_e \approx 2.0$ is the electron g-factor, $\mu_B$ is the Bohr magneton and $S_z$ is the projection of the total spin $S$ on the symmetry axis of the defect (in case of the defects considered here it coincides with the $c$-axis of SiC). Without external magnetic field ($B = 0$) the ground state is split due to the crystal field $D$ and for $S = 3/2$ the zero-field splitting  (ZFS) between the $m_s = \pm 1/2$ and $m_s = \pm 3/2$ sublevels is equal to $2 D$ [the inset of Fig.~\ref{fig1}(b)]. Remarkably, the zero-field splitting is an individual fingerprint for each defect. 
In table~\ref{ZFS} we summarize the ZFS values 
for different intrinsic defects in the most commonly encountered polytype 6H-SiC.  

\begin{table}[btph]
\caption{Zero-field splitting (spin resonance frequency) of intrinsic defects in 6H-SiC.}
\begin{center}
\begin{tabular}{|c|c|c|c|}
$\,\,\,\,\,\,\,  \mathrm{V_{Si}}$-$\mathrm{V_{C}} \,\,\,\,\,\,$ & $\,\,\,\,\,\, \mathrm{V_{Si}(V2)} \,\,\,\,\,\,$ & $\,\,\,\,\,\, \mathrm{V_{Si}(V3)} \,\,\,\,\,\,$ & $\,\,\,\,\,\, \mathrm{V_{Si}}$-$\mathrm{Si_{i}} \,\,\,\,\,\,$ \\
divacancy  & "hexagonal" & "cubic" & Frenkel pair\\
\hline
$1.28 \, \mathrm{GHz}$  & $127 \, \mathrm{MHz}$ & $27 \, \mathrm{MHz}$ & $408 \, \mathrm{MHz}$ \\
\hline
Ref.~\onlinecite{Baranov:2005em} & \multicolumn{2}{c|}{Refs.~\onlinecite{Sorman:2000ij, Riedel:2012jq} } & Ref.~\onlinecite{vonBardeleben:2000jg} \\
$T= 2$~K & \multicolumn{2}{c|}{$T = 2$, $50$~K} &  $T = 300$~K\\
\end{tabular}
\end{center}
\label{ZFS}
\end{table}

We now demonstrate, using optically detected magnetic resonance (ODMR) technique, that defect spins in 6H-SiC can be initialized and read out at ambient conditions, which is the basis for various sensing applications. Optical excitation with $785 \, \mathrm{nm}$ followed by spin-dependent recombination through the metastable state (MS) results in a spin polarization of the defect ground state [the inset of Fig.~\ref{fig1}(b)]. On the other hand, photoluminescence ($\mathrm{PL}$) rates from the excited state (ES) to the ground state (GS) involving different spin sublevels are different. When a resonance RF equal to the ZFS is applied ($\nu_0 = 2 D / h$ for $S = 3/2$), it induces magnetic dipole transitions between the spin-split sublevels resulting in a change of the photoluminescence intensity ($\mathrm{\Delta PL}$).  Since single photons emitted by a single defect can be detected, this ODMR technique is now a standard method to probe single NV spins in diamond \cite{Gruber:1997gk}, which has the resonance frequency around $\nu_0 = 2.87 \, \mathrm{GHz}$. 

In order to probe spin defects in SiC we have extended the RF spectral range from a few GHz down to a few tens of MHz, usually assigned for nuclear magnetic resonance experiments. Figure~\ref{fig1}(b) shows a typical ODMR spectrum, i.e., relative change of the photoluminescence intensity $\mathrm{\Delta PL / PL}$ as a function of applied RF. Two spin resonances at $\nu_0 = 28 \, \mathrm{MHz}$ and $\nu_0 = 128 \, \mathrm{MHz}$ agree well with ZFS of the silicon vacancy defects $\mathrm{V_{Si}(V3)}$ and $\mathrm{V_{Si}(V2)}$, respectively (table~\ref{ZFS}).  Another spin resonance is observed at $\nu_0 = 367 \, \mathrm{MHz}$
and we ascribe it to the Frenkel pair $\mathrm{V_{Si}}$-$\mathrm{Si_{i}}$ (table ~\ref{ZFS}). A discrepancy of 10\% can be explained by the strong dependence of its ZFS on temperature and laser power, as described later in this work (see also Supplemental Material). Finally, we discuss the origin of the spin resonance at $\nu_0 = 1.03 \, \mathrm{GHz}$. We find that this spin resonance is strongly temperature dependent---$D (T)$ is discussed later in detail---and at $T = 10$~K we measure the zero-field splitting to be $\nu_0 = 1.22 \, \mathrm{GHz}$. This agrees reasonably well with ZFS of the divacancy $\mathrm{V_{Si}}$-$\mathrm{V_{C}}$ (table~\ref{ZFS}). On the other hand, we register PL up to $1050 \, \mathrm{nm}$, but the $\mathrm{V_{Si}}$-$\mathrm{V_{C}}$ PL has maximum above  $1100 \, \mathrm{nm}$ \cite{Son:2006im}. We would like to mention that the magnetic field dependences (presented later) unambiguously indicate $S = 3/2$ for this defect, while in the earlier studies the $S = 1$ ground state of the divacancy was assumed \cite{Baranov:2005em}. For this reason we label it as unidentified defect (UD) in all figures, indicating that its origin should be carefully examined, which is beyond the scope of this work.  

\begin{figure}[btp]
\includegraphics[width=0.99\columnwidth]{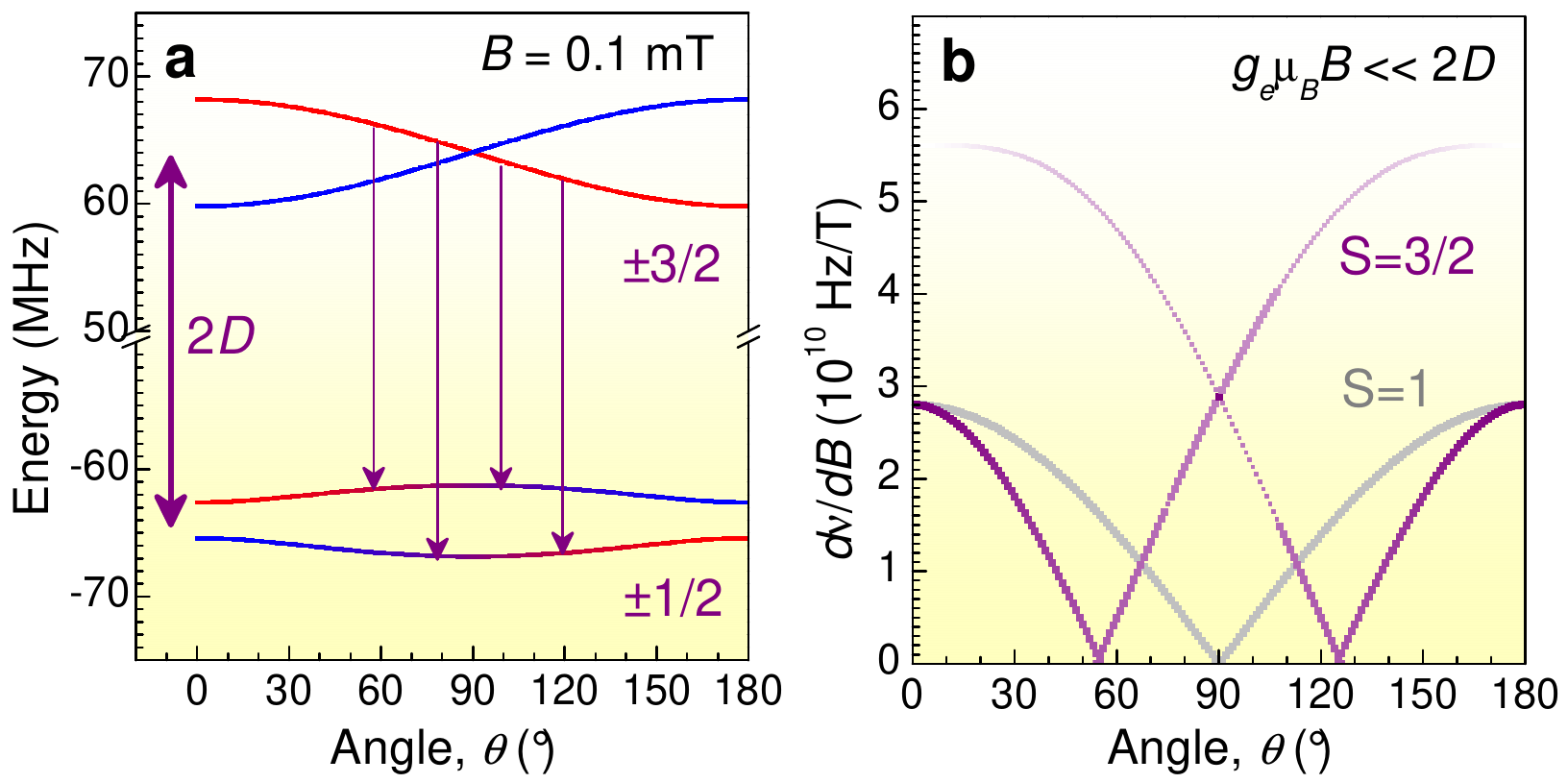}
\caption{Vector magnetometry. (a) Zeeman splitting of the $\pm 3/2$ (upper) and $\pm 1/2$ (lower) spin sublevels as a function of the angle $\theta$ between magnetic field $B = 0.1 \, \mathrm{mT}$ and the defect symmetry axis ($2 D = 128 \, \mathrm{MHz}$). The arrows indicate four possible RF-induced transitions. (b) Comparison of the relative shifts of the spin resonance frequencies $d \nu / d B$ as a function of $\theta$ for $S=1$ and $S=3/2$.  } \label{fig2}
\end{figure}

First, as a characteristic example, we consider theoretically the Zeeman splitting of the $\mathrm{V_{Si}(V2)}$ spin sublevels for  different orientations (given by the polar angle $\theta$) of a weak magnetic field  $B \ll 2 D / g_e \mu_B$ [Fig.~\ref{fig2}(a)]. The $\mathrm{V_{Si}(V2)}$ defect has spin $S = 3/2$ ground state \cite{Wimbauer:1997fj, Mizuochi:2002kl, Kraus:2013di} and the upper $m_s = \pm 3/2$ spin sublevels are split as $\pm \frac{3}{2}  g_e \mu_B B \cos \theta $. Interestingly, the lower $m_s = \pm 1/2$ spin sublevels are mixed due to the perpendicular-to-the-symmetry-axis field component and split differently as $\pm  \frac{1}{2} g_e \mu_B B (1 +  3 \sin^2 \theta)^{1/2}$, which results in four possible RF-induced transitions.  We obtain the probability of these transitions [color-coded in Fig.~\ref{fig2}(b)] as 
\begin{equation}
W_{jk} \sim  \big| \langle j | \mathbf{B_1 S} | k \rangle \big|^2\,,
 \label{Transitions}
\end{equation}
where $| j \rangle$, $| k \rangle$ are the eigenstates of (\ref{Hamiltonian}) and $B_1 \bot c$ is the driving RF field. 

Figure~\ref{fig2}(b) compares the relative shifts of the spin resonance frequency $d \nu / d B$ depending on the magnetic field orientation for $S = 1$ and $S = 3/2$. In case of a spin-1 system (i.e., the NV defect) it follows $| g_e \mu_B  \cos \theta / h |$ ($h$ is the Planck constant) and tends to zero for $B \bot c$ ($\theta = 90 ^\circ$). Therefore, using a single (or equally oriented) spin-1 defect(s) the field projection on the defect symmetry axis $B_z = B  \cos \theta$ rather than the absolute value is detected.  In fact, due to the presence of the nuclear spin bath, the NV spin-echo decay curve contains characteristic "collapses and revivals" from which the total magnetic field can be determined \cite{Childress:2006km}. However, the nuclear gyromagnetic ratio is much smaller (by a factor 0.0003 for $\mathrm{^{13}C}$) than that of electron, making this method less sensitive. Furthermore, this method is not applicable for isotopically purified crystals, providing the highest sensitivity. In contrast, in case of a spin-3/2 system there are two pairs of resonances. 
Therefore, from the comparison of the relative ODMR frequencies and amplitudes both the absolute value $B$ and its orientation with respect to the defect symmetry axis can be reconstructed. 

\begin{figure}[btp]
\includegraphics[width=0.95\columnwidth]{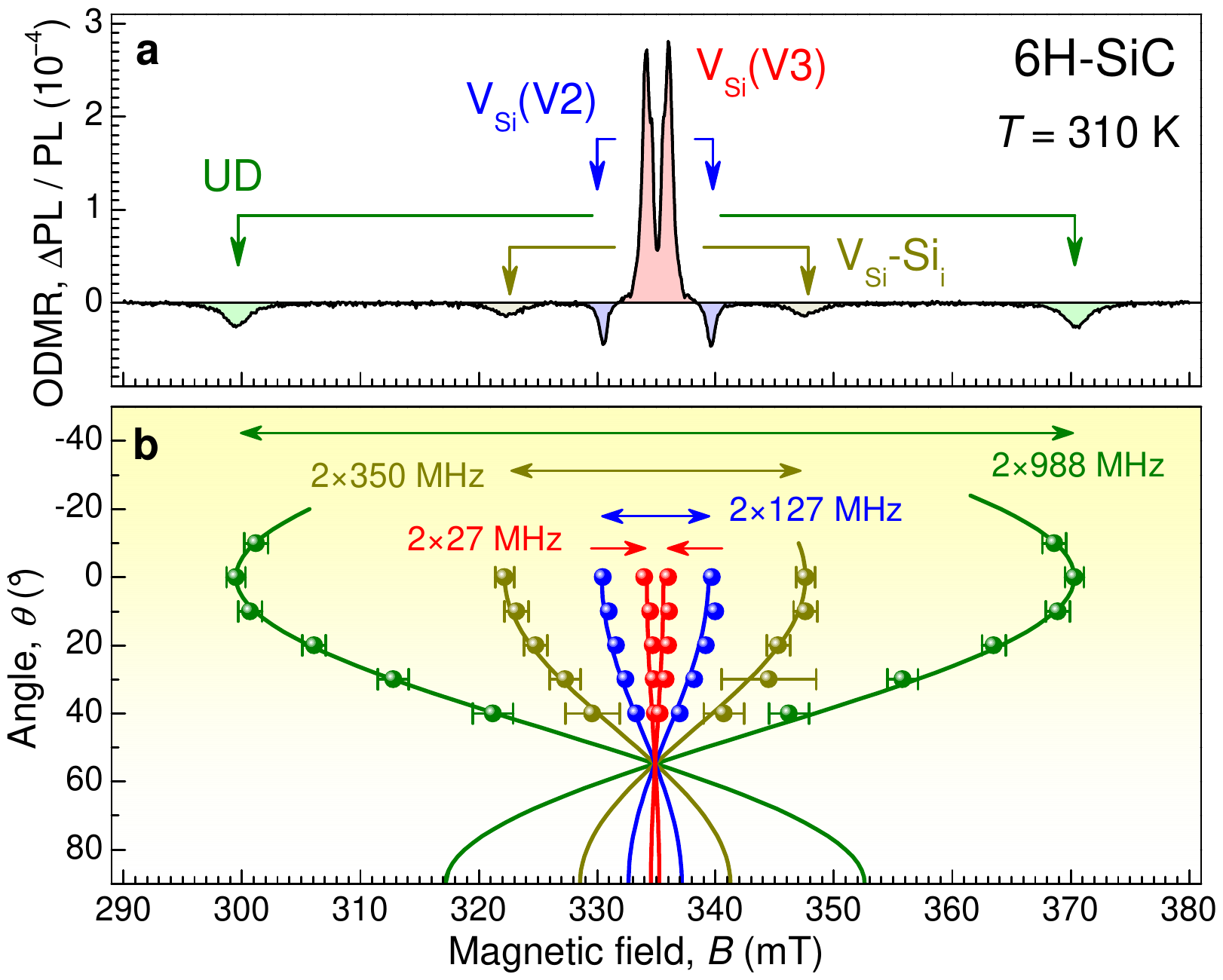}
\caption{Axial symmetry of spin defects in 6H-SiC. (a) Room-temperature ODMR spectrum in the X-band ($\nu = 9.4 \, \mathrm{GHz}$) as a function of  magnetic field $B | | c$ . (b) ODMR frequencies for different angles between the magnetic field $B$ and the $c$-axis of 6H-SiC.  Experimental data are shown by symbols, the solid lines correspond to the calculations to Eq~(\ref{AngleH}). } \label{fig3}
\end{figure}

In order to demonstrate this advantage of the spin-3/2 defects in SiC for magnetic sensing applications it is necessary to ascertain their symmetry axes. This can be done in high magnetic fields  $B \gg 2 D / g_e \mu_B$. In this case a pair of spin resonances should be observed for each spin defect, the separation between those being a function of $\theta$. We have performed this experiment in the X-Band at a fixed frequency  $\nu = 9.4 \, \mathrm{GHz}$ while sweeping the magnetic field and an ODMR spectrum obtained for $B | | c$ is shown in Fig.~\ref{fig3}(a). Indeed, four pairs of spin resonances from four distinct spin defects are observed, in accord with four lines in the zero-field ODMR spectrum of Fig.~\ref{fig1}(b). The resonance magnetic fields $B_{\pm}$ related to each pair are presented in Fig.~\ref{fig3}(b) as a function  of the angle between the field direction and the $c$-axis. 

According to spin Hamiltonian~(\ref{Hamiltonian}), the difference between these fields is given by 
\begin{equation}
 \Delta B =   B_+ - B_- = \frac{2 D (T)  }{ g_e \mu_B }  (3 \cos^2 \theta - 1)   \,.
 \label{AngleH}
\end{equation}
As Eq~(\ref{AngleH}) describes the experimental data quite well [Fig.~\ref{fig3}(b)], one immediately concludes that all the defects under consideration are in line with the $c$-axis, as also depicted in Fig.~\ref{fig1}(a). Furthermore, for  $B | | c$ ($\theta = 0 ^\circ$) ZFS can independently be measured \cite{Riedel:2012jq} as $2 D = g_e \mu_ B \Delta B /2$ and from the comparison with the zero-field data of Fig.~\ref{fig1}(b) all the resonances in Fig.~\ref{fig3} are unambiguously assigned.  

\begin{figure}[btp]
\includegraphics[width=0.95\columnwidth]{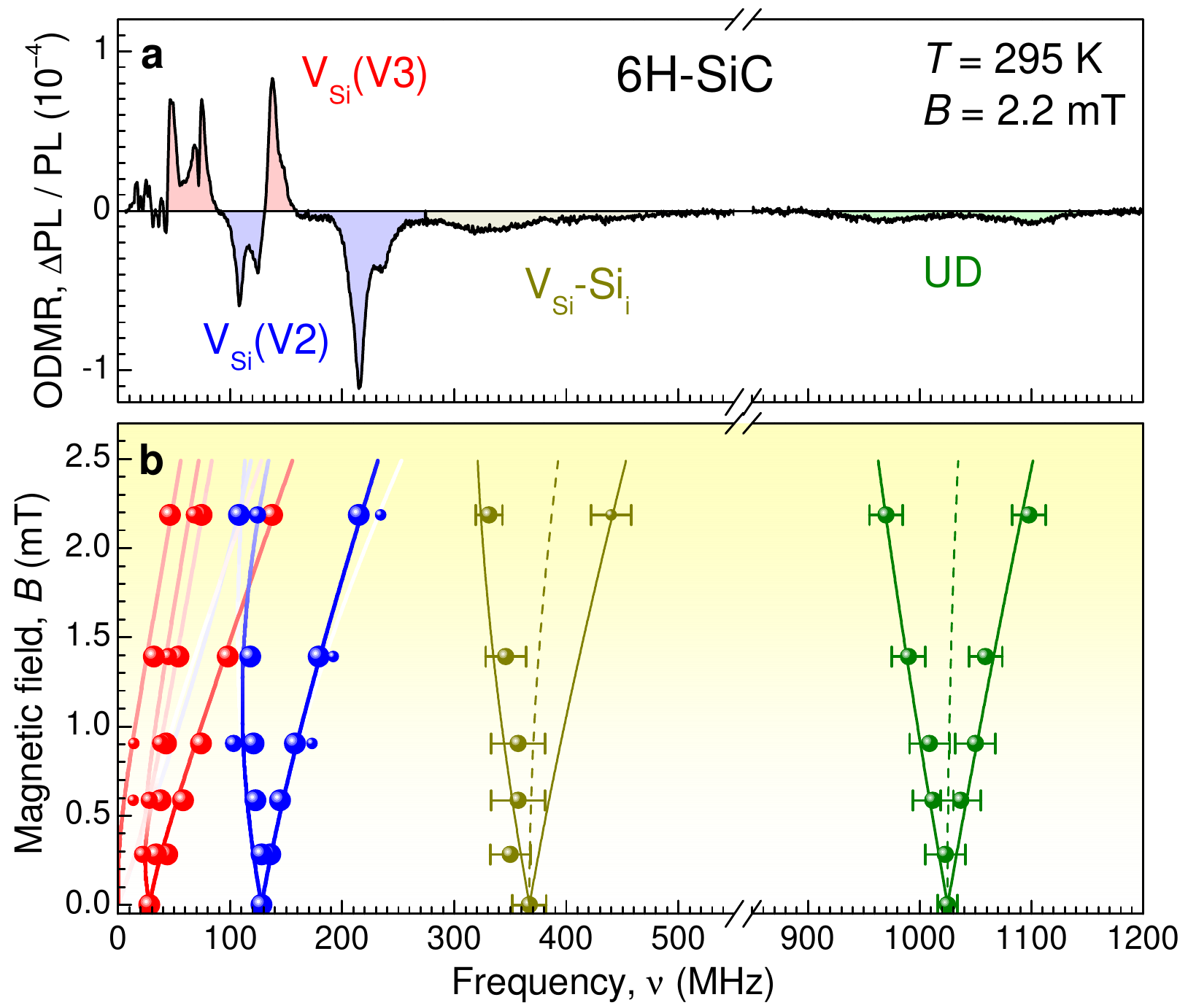}
\caption{Effect of external magnetic field in 6H-SiC. (a) Room-temperature ODMR spectrum obtained in a magnetic field $B =  2.2 \, \mathrm{mT}$. (b) ODMR frequencies as a function of magnetic field. Experimental data are shown by symbols, the solid and dotted lines correspond to the calculations for $S = 3 / 2$ and  $S = 1$, respectively. In all panels $\mathbf{B} \bot c$.} \label{fig4}
\end{figure}

Having established the orientation of the spin defects in our 6H-SiC sample, we apply a weak magnetic field $B = 2.2 \, \mathrm{mT}$ perpendicular to the defect symmetry axis ($\theta = 90 ^\circ$)  [Fig.~\ref{fig4}(a)]. In this configuration the difference between different spin systems is most pronounced [see Fig.~\ref{fig2}(b)]. For a spin-1 defect such oriented weak magnetic fields should have no effect. In contrast, we observed large Zeeman splittings for all four ODMR lines [symbols Fig.~\ref{fig4}(b)]. The experimental data are perfectly described by the spin Hamiltonian~(\ref{Hamiltonian})  with $S=3/2$ and corresponding transition rates of Eq.~(\ref{Transitions}), as shown by the solid lines in Fig.~\ref{fig4}(b). For comparison, the calculations with $S=1$ are shown by the dotted lines. Remarkably, for $\mathrm{V_{Si}(V2)}$ and especially for $\mathrm{V_{Si}(V3)}$ the magnetic field behavior is not linear because the Zeeman splitting in a magnetic field of a few mT is comparable or larger than ZFS in these silicon vacancy defects, which results in the intermixing of all four spin sublevels. According to the magnetic field dependencies of Fig.~\ref{fig4}(b), all defects under consideration have the $S =3/2$ ground state. This is in agreement with the earlier ascertained  spin structure of the $\mathrm{V_{Si}}$ defects \cite{Kraus:2013di} and with the model of the $\mathrm{V_{Si}}$-$\mathrm{Si_{i}}$ Frenkel pairs \cite{vonBardeleben:2000jg}. 

We now discuss the effect of strain. In case of $S = 1$ spin system, as the NV defect in diamond and the divacancy in SiC,  the local strain lifts up the degeneracy of the $m_s = \pm 1$ spin sublevels even without external magnetic field. It is described by an additional term $E (S_x^2 - S_y^2)$ in the spin Hamiltonian, where $E$ is the transverse ZFS parameter. This may limit the magnetometry sensitivity because the spin-splitting in external magnetic field for $E \neq 0$ is described by $\sqrt{(g_e \mu_B B_z)^2 + E^2}$, i.e., the relative shift of the spin resonance frequency is quadratic rather than linear for small $B$. By improving the quality of bulk diamond material, the parameter $E$ can be reduced below $1 \, \mathrm{MHz}$ and further eliminated in a small bias magnetic field. However, magnetic fields may lead to additional inhomogeneity, which is in some experiments rather unwanted. Furthermore, in nanodiamonds, which are very interesting for bio-sensing, strain is always present and the transverse ZFS parameter can be as large as $E =  20 \, \mathrm{MHz}$ \cite{Bradac:2010bm}. The situation is qualitatively different for half-integer spin systems, like spin defects with $S = 3 / 2$. According to the Kramers theorem, the double degeneracy of the $m_s = \pm 1 / 2$ and $m_s = \pm 3 / 2$ states can only be lifted up by an external magnetic field, meaning that essentially $E = 0$. This makes these defects (in particular, silicon vacancies in SiC) robuster against strain fluctuations.  

\begin{figure}[btp]
\includegraphics[width=0.95\columnwidth]{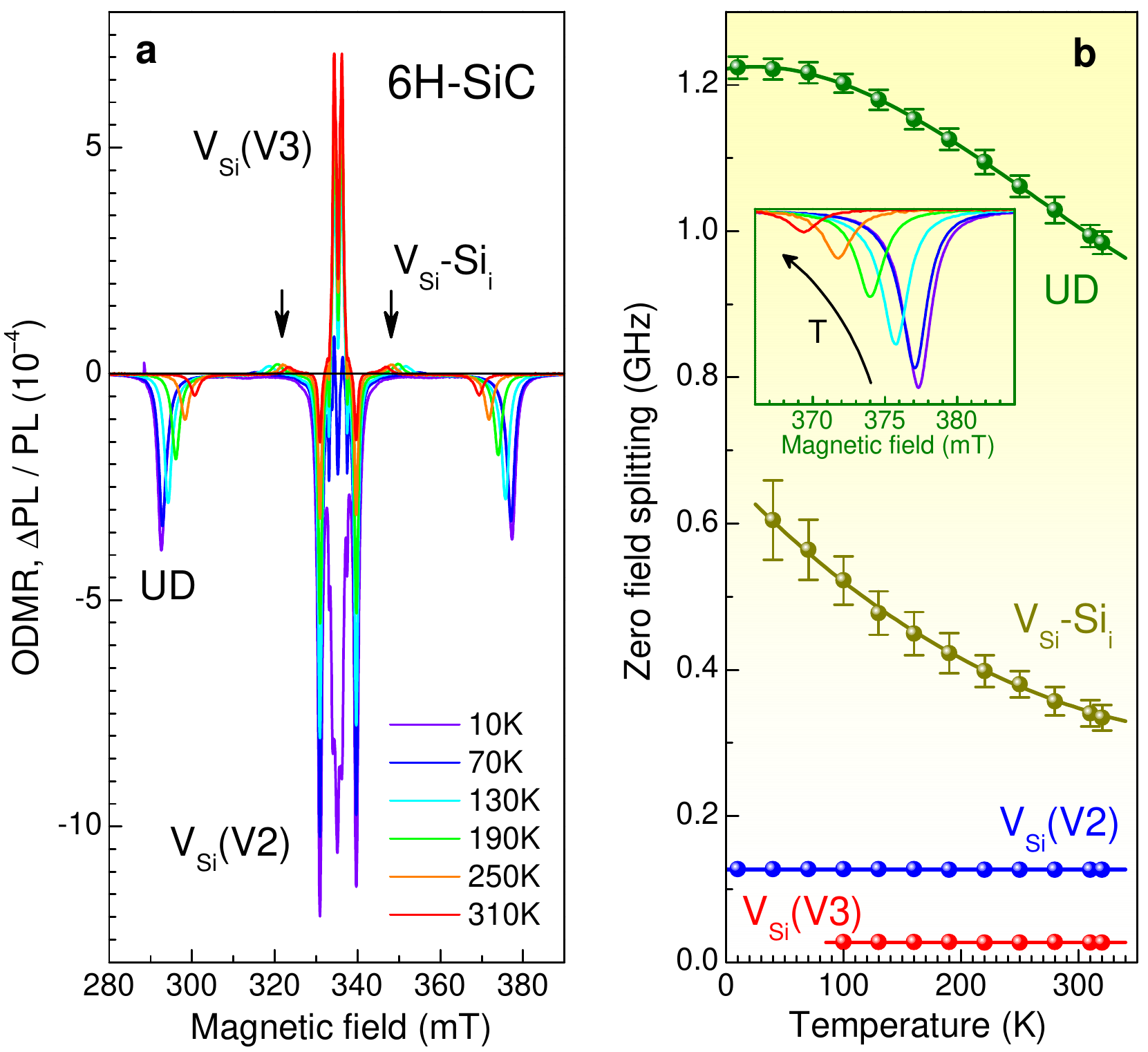}
\caption{Temperature effect in 6H-SiC. (a) ODMR spectra in the X-band ($\nu = 9.4 \, \mathrm{GHz}$) recorded at different temperatures. (b) Zero-field splitting $2 D$ for different spin defects as a function of temperature. The solid lines are third order polynomial fits. Inset: The same as (a), but rescaled to highlight a strong UD temperature dependence.} \label{fig5}
\end{figure}

In oder to examine the effect of temperature fluctuations, we have measured ODMR spectra in the temperature range from $10$ to $320 \, \mathrm{K}$ [Fig.~\ref{fig5}(a)]. The experiment is performed in the X-band and $D$ for each defect is determined using Eq.~(\ref{AngleH}). As one can see from Fig.~\ref{fig5}(b) the parameter $D$ of the $\mathrm{V_{Si}(V2)}$ and $\mathrm{V_{Si}(V3)}$ defects is temperature independent within the accuracy of our experiment (a few $\mathrm{kHz / K}$). This is an additional strong argument to use these defects for magnetometry. 

In contrast to $\mathrm{V_{Si}(V2)}$ and $\mathrm{V_{Si}(V3)}$, for the Frenkel pair $\mathrm{V_{Si}}$-$\mathrm{Si_{i}}$ we observe a reduction of its ZFS ca. 50\%, from $0.6$ down to $0.3  \, \mathrm{GHz}$ when the temperature increases from $10$ to $320 \, \mathrm{K}$ [Fig.~\ref{fig5}(b)].  The corresponding temperature dependence is well described by a third order polynomial $2 D (T) = a_0 + a_1 T + a_2 T^2 + a_3 T^3$ with $a_0 = (667 \pm 4) \, \mathrm{MHz}$, $a_1 = (- 1.7 \pm 0.1) \, \mathrm{MHz / K}$, $a_2 = (2.3 \pm 0.8) \times 10^{-3} \, \mathrm{MHz / K^2}$, and $a_3 = (-8 \pm 15) \times 10^{-7} \, \mathrm{MHz / K^3}$. The defect with the largest ZFS (labeled as UD in Fig.~\ref{fig5}) has a slightly weaker temperature dependence (17\% relative reduction of $D$) but reveals the strongest thermal shift $2 d D / d T$  at room temperature [the inset of Fig.~\ref{fig5}(b)]. From the polynomial fit with $a_0 = (1222 \pm 2) \, \mathrm{MHz}$, $a_1 = (2.4 \pm 0.4) \times 10^{-1} \, \mathrm{MHz / K}$, $a_2 = (- 5.2 \pm 0.3) \times 10^{-3} \, \mathrm{MHz / K^2}$, and $a_3 = (6.5 \pm 0.5) \times 10^{-6} \, \mathrm{MHz / K^3}$ we find $\beta = 2 d D / d T = -1.1 \, \mathrm{MHz / K}$ at $T = 300 \, \mathrm{K}$. This value is 14 times larger than the thermal shift of the NV defect in diamond \cite{Acosta:2010fq}.

\section{Intrinsic defects in 4H-SiC and magnetic field sensing}

\begin{figure}[tpb]
\centerline{\includegraphics[width=0.95\columnwidth]{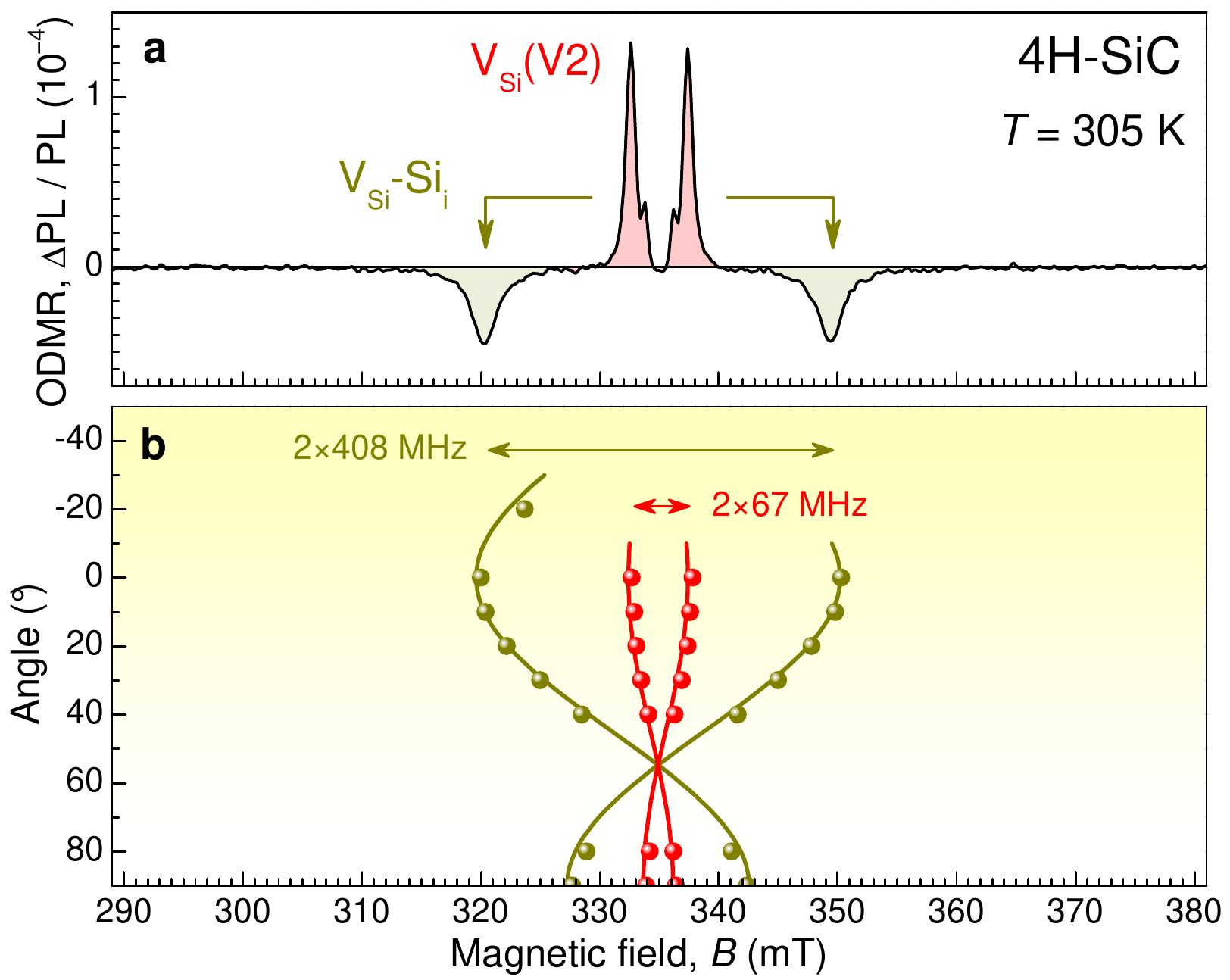}}
\caption{Room-temperature ODMR in 4H-SiC. (a) An ODMR spectrum in the X-band ($\nu = 9.4 \, \mathrm{GHz}$) as a function of  magnetic field $\mathbf{B} | | c$. (b) ODMR frequencies for different angles between the magnetic field $B$ and the $c$-axis.  Experimental data are shown by symbols, the solid lines correspond to the calculations to Eq~(\ref{AngleH}).} \label{fig6}
\end{figure}

We also observe very similar behaviour in 4H polytype of SiC, which can be grown of high quality. Figure~\ref{fig6}(a) shows a room-temperature ODMR spectrum of 4H-SiC, obtained in the X-Band spectrometer. Following the angle dependencies presented in Fig.~\ref{fig6}(b) we ascribe the ODMR lines to the  silicon vacancy $\mathrm{V_{Si}(V2)}$ (with ZFS of 67~MHz) and the Frenkel pair $\mathrm{V_{Si}}$-$\mathrm{Si_{i}}$ (with ZFS of 408~MHz) \cite{Sorman:2000ij, Riedel:2012jq, vonBardeleben:2000jg}. 

\begin{figure}[tpb]
\centerline{\includegraphics[width=0.99\columnwidth]{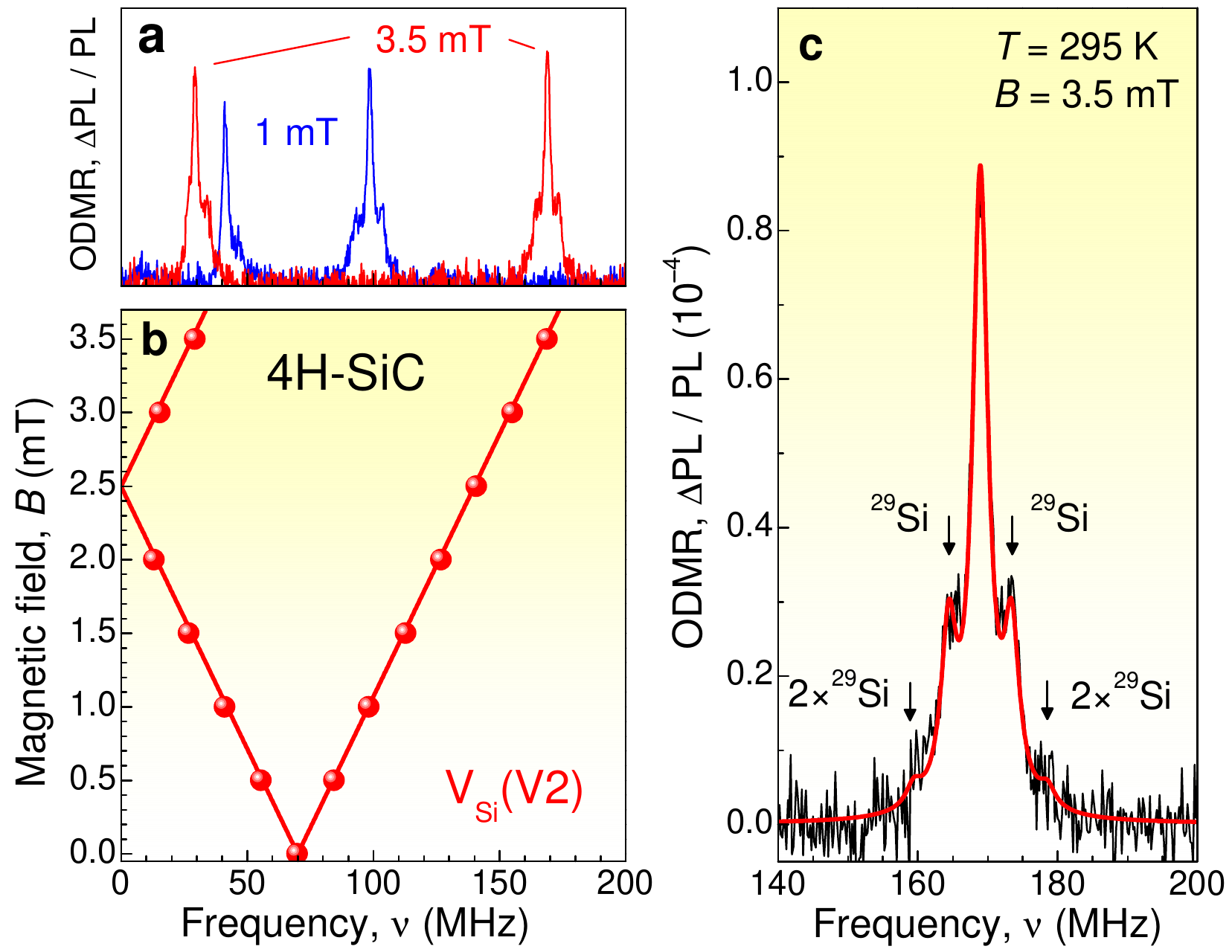}}
\caption{Magnetic field sensing. (a) Room-temperature ODMR spectra of 4H-SiC obtained in magnetic fields $B =  1 \, \mathrm{mT}$ and $B =  3.5 \, \mathrm{mT}$. (b) The $\mathrm{V_{Si}(V2)}$ ODMR frequencies as a function of magnetic field for $\mathbf{B} | | c$. Experimental data are shown by symbols, the solid lines correspond to the calculation as explained in the text. (c)  An ODMR line of $\mathrm{V_{Si}(V2)}$ fitted to Eq~(\ref{LineShape}). Arrows indicate the positions of the $\mathrm{^{29}Si}$ isotope peaks. } \label{fig7}
\end{figure}

We now concentrate on the ODMR spectra of the $\mathrm{V_{Si}(V2)}$  defect in zero and weak bias magnetic fields [Fig.~\ref{fig7}(a)]. In zero magnetic field, the ODMR line has maximum around $\nu_0 \approx 70 \, \mathrm{MHz}$ [Fig.~\ref{fig7}(b)]. Upon application of a magnetic field along the $c$-axis ($\theta = 0 ^\circ$), the ODMR line is split and one observes two resonances [Fig.~\ref{fig7}(a)]. They shift linear with magnetic field as $\nu_B = | \nu_0  \pm g_e \mu_B  B/ h |$ [Fig.~\ref{fig7}(b)], in accord to Hamiltonian~(\ref{Hamiltonian}) for $\mathbf{B} | | c$. From the linear fit we find the $\mathrm{V_{Si}(V2)}$ ZFS in our crystal with high accuracy $\nu_0 = 70.2  \pm 0.3 \, \mathrm{MHz}$. 

In our earlier studies we have reported that the ODMR linewidth in SiC strongly depends on the RF power \cite{Kraus:2013di}. Therefore, to achieve the narrowest linewidth we use the lowest RF power, which still provides a reasonable signal-to-noise ratio. Additionally, we apply a small bias magnetic field of $3.5 \, \mathrm{mT}$ to eliminate the broadening  due to environment magnetic fields. The result is presented in Fig.~\ref{fig7}(c) and can be well fitted to
\begin{equation}
 \frac{ \mathrm{\Delta PL}  }{ \mathrm{PL} } (\nu) =   \sum_{n = 0,1,2 }  \frac{\Gamma_n}{ [2 \pi T_2^* (\nu_B \pm n A_{\mathrm{Si}} /2 - \nu)]^2 +1 }\,.
 \label{LineShape}
\end{equation}
Here, $A_{\mathrm{Si}}$ is the hyperfine constant describing the interaction $\mathcal{H} = \mathbf{I} A \mathbf{S}$ with the next-nearest-neighbor (NNN) $\mathrm{^{29}Si}$ ($I = 1/2$) nuclear spins. We find $A_{\mathrm{Si}} = 9.1 \pm 0.2  \, \mathrm{MHz}$, which is comparble with the earlier reported value \cite{Sorman:2000ij}. The relative amplitudes of the peaks $\Gamma_0 : \Gamma_1 : \Gamma_2$ are given by the probability to find $n$ spin-active $\mathrm{^{29}Si}$ isotopes among the 12 Si NNN. We find $ 0.89 \times 10^{-4} : 0.24 \times 10^{-4} : 0.36 \times 10^{-5}$, which is close to the ratios expected for the natural abundance of $\mathrm{^{29}Si}$ (4.7\%). 

The resonance linewidth in Fig.~\ref{fig7}(c) is inversely proportional to the inhomogeneous spin coherence time, which is $T_2^* = 120 \, \mathrm{ns}$ according fit to Eq.~(\ref{LineShape}). The accuracy to measure the relative change of the frequency ($\delta \nu$) is given by the maximum value of the derivative $(\delta \Gamma / \delta \nu )_{max} =  \pi T_2^* \Gamma_0$. With the root-mean-square deviation $\delta \Gamma = 3.5 \times 10^{-6}$ [as in Fig.~\ref{fig7}(c)] we obtain $\delta \nu = 100 \, \mathrm{kHz}$. Taking into account the integration time of $8 \, \mathrm{s}$ per point, this corresponds to the magnetic field sensitivity $\delta B = 10 \, \mathrm{ \mu T  /  \sqrt{Hz}}$. The sensitivity can be improved by increasing of $T_2$. It is limited by the spin-lattice relaxation time, which is $T_1 = 100 \, \mathrm{\mu s}$ \cite{Soltamov:2012ey} for $\mathrm{V_{Si}}$ in SiC and can be achieved using spin-echo techniques. Further improvement by an approximately an order of magnitude is also possible in isotopically purified $\mathrm{^{28}Si^{12}C}$, as was proposed recently \cite{Riedel:2012jq}. For instance, the record spin coherence time of the NV defect in $\mathrm{^{12}C}$-diamond is $T_2 = 1.8 \, \mathrm{ms}$ \cite{Balasubramanian:2009fu}. In order to compare the sensitivity of the $\mathrm{V_{Si}}$  defect with the sensitivity of the NV defect, we first estimate the product of the overall detection efficiency ($\mathcal{C}$) and the number of the $\mathrm{V_{Si}(V2)}$ defects ($N$) in our experiments, by taking into account the signal level (ca. $10 \, \mathrm{nW}$) and the PL lifetime ($6 \, \mathrm{ns}$, ref.~\onlinecite{Hain:2014tl}). We then estimate the expected sensitivity for optimized samples to be on the order of magnitude $\delta B = 10 \, \mathrm{ nT } /  \sqrt{ \mathcal{C} N \cdot \mathrm{ Hz }}$. 

Our experiments have been performed on an ensemble of defects. Very recently, the detection of some single defect centers in SiC has been reported \cite{Castelletto:2013el}, and we believe that the same approach can be used to isolate single $\mathrm{V_{Si}}$-related defects. An important issue in this case is the ODMR contrast. It increases with RF power and we observe $\Gamma_0 = 0.11 \%$ for RF power of $36 \, \mathrm{dBm}$. 
This may lead to some limitations compared to single NV defects with larger ODMR contrast ($\Gamma_0 \sim 10\%$), such as a requirement of longer integration time to achieve the equal sensitivity with the same other parameters. On the other hand, we collect PL from various types of defects and only one of them gives rise to the ODMR contrast. This means that $\Gamma_0$ of a single defect should be larger than in ensemble experiments. Alternatively, the ODMR contrast can be sizeably increased by a proper choice of the spectral detection window \cite{Falk:2013jq}. Furthermore, in many applications it is not necessary to use just one defect. 
All the silicon vacancy related defects considered here are aligned along the same axis, and, as intrinsic defects, they can be created of high concentration without need for additional doping.

\section{Temperature sensing}

We now discuss the temperature sensitivity, which is proportional to $(2 d D / d T)^{-1} \delta \nu$ and therefore it is natural to select a defect with a large thermal shift $\beta$, such as the Frenkel pair $\mathrm{V_{Si}}$-$\mathrm{Si_{i}}$ [Fig.~\ref{fig5}(b)]. A large change of ZFS with temperature cannot be explained simply by thermal lattice expansion. Assuming that $D$ is due to dipolar coupling $\propto r^{-3}$ between the silicon vacancy and interstitial Si atom ($r$ is the distance between them),  one should have $d D / (D d T) \approx -3 \alpha$. Here, $\alpha = 4.0 \times 10^{-6}  \, \mathrm{K^{-1}}$ is the thermal expansion coefficient of SiC, but we observe two orders of magnitude larger value of  $2  \times 10^{-3}  \, \mathrm{K^{-1}}$. A possible explanation is that the Si interstitial ($\mathrm{Si_{i}}$) is not rigid in the SiC lattice and a small perturbation may result in a significant shift in position relative to $\mathrm{V_{Si}}$, leading to a large change of ZFS. 

The UD defect demonstrates even larger thermal shift. Using the same procedure as for the $\mathrm{V_{Si}(V2)}$ defect we estimate the temperature sensitivity ca. $\delta T = 1 \, \mathrm{ K  /  \sqrt{Hz}}$. Remarkably, the ODMR linewidth is significantly larger than for $\mathrm{V_{Si}(V2)}$  and can be explained by a high sensitivity to the local environment. In high-quality samples or for single defects the inhomogeneous broadening should be suppressed, leading to homogenous linewidth $1 / (\pi T_2)$.  In isotopically engineered samples and with the  use of  advanced measurement protocols \cite{Toyli:2013cn, Neumann:2013hc, Kucsko:2013gq, Fang:2013dw} the projected temperature sensitivity is $\delta T = 1 \, \mathrm{ mK } /  \sqrt{ \mathcal{C} N \cdot \mathrm{ Hz }}$. 

Another potential application could be a synchronized composite clock, as was  originally proposed for the NV defects in diamond \cite{Hodges:2013cs}. We find the Allan variance $ (\delta \Gamma / \Gamma_0) / (2 \pi Q  \sqrt{\tau})$  corresponding to the resonance in  Fig.~\ref{fig7}(b) to be  $2 \times 10^{-4} / \sqrt{\tau}$. By applying a spin-echo technique to isotopically purified samples it should be possible to significantly increase the quality factor $Q = \pi T_2 \nu_0$ and with  $T_2 = 1 \, \mathrm{ms}$ the Allan variance approaches  $10^{-8} / \sqrt{\tau}$. Further improvement by several orders of magnitude can potentially be achieved by increasing the number of defects (signal-to-noise ratio $\Gamma_0 / \delta \Gamma \propto \sqrt{N}$), as this does not require intentional doping with paramagnetic impurities (i.e., nitrogen for the NV defect in diamond) and hence is not accompanied by dephasing effects. In this case, the main limiting factor is temperature uncertainty. For instance, the temperature stabilization within $\Delta T =  0.01 \, \mathrm{K}$ leads to a fractional frequency stability $\Delta T  \beta / \nu_0 < 10^{-6}$ for the $\mathrm{V_{Si}}$ defects in SiC.  To compensate temperature fluctuations, it has been proposed to lock the frequency difference between two defects with different thermal shifts ($\beta_1$ and $\beta_2$), resulting in a frequency uncertainty scaled by a factor $\sqrt{ 1+ 2(\beta_1 / \Delta \beta_{1,2})^2}$ relative to the temperature-insensitive case \cite{Hodges:2013cs}. This idea can be implemented within one SiC crystal by using different types of defects. For instance, the use of  the $\mathrm{V_{Si}(V2)}$ defect ($ \beta_1$) and the UD defect ($\beta_2$) in 6H-SiC yields $\beta_1 / \Delta \beta_{1,2} < 10^{-2}$.

Summarizing, intrinsic defects in SiC demonstrate complementary characteristics to the NV defect in diamond. The variety of defect's properies makes them very attractive for various sensing applications, such as magnetometry, thermometry and chip-scale timekeeping. An important aspect is the possibility for integration and control of these defects within SiC-based electronic devices \cite{Fuchs:2013dz}. Another intriguing possibility is to use biocompatibility of SiC nanocrystals to perform nanonscale magnetic field and temperature imaging in biological systems, which can be done at a fixed RF frequency  as well \cite{Babunts:2012bu}.

\section{Methods summary}

We present experimental results for 6H and 4H polytype SiC crystals grown by the standard sublimation technique. In order to generate defects our crystals have been irradiated with neutrons ($5 \, \mathrm{MeV}$) to a dose of $10^{16} \, \mathrm{cm^{-2}}$. 

The full-field ODMR experiments are performed in a X-band cavity ($Q = 3000$, $\nu = 9.4$~GHz) with direct optical access. The samples are mounted in a liquid helium flow cryostat with a possibility to vary temperature from 10~K to slightly above room temperature. The temperature sensor is situated in the cryostat about 1~cm below the sample. A diode laser operating at 785~nm (100~mW) is used to optically pump all types of defects in SiC. The PL excited by this laser is passed through a 950-nm longpass filter and detected by a Si photodiode (up to $1050 \, \mathrm{nm}$).

In the zero-field ODMR experiments, the RF radiation provided by a signal generator is amplified up to 36~dBm (4~W) and guided to a thin copper wire terminated with 50-$\Omega$  impedance. The laser beam (15~mW) is focused close to the wire using an optical objective. The filtered PL signal (10~nW) is coupled to an optical fiber and detected by a Si photodiode. Weak magnetic fields are applied using a permanent 
magnet, the field orientation and strength being ascertained using ODMR on the NV defects in diamond as a reference. In both experiments we chop the RF radiation  (typically below $1 \, \mathrm{kHz}$) and the output PL signal is locked-in.


\section*{Acknowledgments}

This work has been supported by the Bavarian Ministry of Economic Affairs, Infrastructure, Transport and Technology, Germany as well as by the DFG under grant AS310/4.


\end{document}